# Ab initio study of CrNF: the first half-metallic ferromagnet nitride fluoride mimicking $CrO_2$

Samir F. Matar[a,*]

*CNRS, ICMCB, Université de Bordeaux, 87 Avenue du Docteur Albert Schweitzer, 33600 Pessac, France*
E-mail: matar@icmcb-bordeaux.cnrs.fr

\* Correspondence and reprints: CNRS, Université de Bordeaux, ICMCB, 87 Avenue du Docteur Albert Schweitzer, 33600 Pessac, France, Tel.: +33–5–4000–2690; Fax: +33–5–4000–2761

*E-mail address*: matar@icmcb-bordeaux.cnrs.fr (S. F. Matar)

*A B S T R A C T*

Based on the different covalent versus ionic chemical bonding and isoelectronic rutile $CrO_2$, CrNF is proposed starting from ordered rutile derivative structures subjected to full geometry optimizations. The ground state structure defined from cohesive energies is of $MgUO_4$-type, characterized by short covalent Cr-N and long ionic Cr-F distances. Like $CrO_2$ it is a half-metallic ferromagnet with M = 2 $\mu_B$ integer magnetization per formula unit with reduced band gap at minority spins. Major differences of magnetic response to pressure characterize CrNF as a soft ferromagnet versus hard magnetic $CrO_2$. The chemical bonding properties point to prevailing covalent Cr-N versus ionic Cr-F bonding. Synthesis routes with two different protocols are proposed and analyzed.

# 1. Introduction

The search for synthetic novel materials with well targeted physical properties is a strategic route for proposing solutions in the field of Materials Science. Simple rules can be followed to devise new stoichiometries with specific properties. As an example boron nitride is an artificial binary which 'mimics' carbon: From the electronic count standpoint, the configurations of the neutral atoms: B ([He] $2s^2\ 2p^1$), N ([He] $2s^2\ 2p^3$) and C ([He] $2s^2 2p^2$), are such that "B + N = 2C", i.e. summing up to 8 valence electrons on each side of the equality sign. Cubic boron nitride, $c$-BN is a good substitute for diamond with a hardness magnitude just below 10 which stands for the value assigned to diamond on the Moh scale of hardness. This property is partly because, unlike diamond, BN does not dissolve into iron, nickel, and related alloys at high temperatures in tooling machines. Due to this electronic similarity, BN adopts similar atomic arrangements as in carbon: three-dimensional (3D) diamond ($c$-BN) as well as two-dimensional (2D) graphite ($h$-BN). Also many other compositions were devised artificially within the BCN diagram such as ultra-hard $C_3N_4$ based on quantum mechanical calculations and respecting the isoelectronic rule (for a review cf. ref. [1] and therein cited works).

Consequently beside structural characteristics, the way the same number of electrons is (re)distributed among chemically different atoms engaged in different interatomic bonds can be of relevance to predict new stoichiometries and properties. Considering an anionic sublattice for a given cation, the isoelectronic relationship for valence shell states is established: O + O → N + F, i.e. 2 x ($2s^2$, $2p^4$) → ($2s^2$, $2p^3$) + ($2s^2$, $2p^5$). Nitride-fluorides of formulation $A^{IV}$NF can be considered as pseudo-oxides, isoelectronic of $A^{IV}O_2$ where $A^{IV}$ is a generic tetravalent element. As simple chemical rules predict, $A$ forms a covalent bond with nitrogen, while an ionic bond is expected to form with fluorine. This chemical bonding difference is likely to generate original physical properties versus oxides as shown in this work. Several nitride-fluorides exist, such as alkaline-earth based $Ba_2NF$ [2] and $Mg_2NF$ [3] and transition metal based TiNF [4] as well as ZrNF [5]. Their investigations carried out within the well established quantum mechanical density functional theory DFT [6,7] led to provide a better understanding of such mixed-anions compounds with potential properties like anisotropic electric conduction, optical anisotropy as well as its growth in thin films or at nanoscopic scales [8]. In TiNF [9] and ThNF [10] ordering between N and F sublattices was analyzed through DFT computations focusing on the energetic and chemical bonding and group-subgroup relations.

$CrO_2$, one of the rare ferromagnetic oxides at room temperature ($T_C \sim 400$ K) and 2 $\mu_B$ magnetization implicitly assigned to tetravalent Cr with $3d^2$ configuration, is known for its use in magnetic recording applications [11]. Since its early discovery [12], several works were devoted to its physical properties [13]. DFT band structure calculations go beyond the phenomenological description by identifying a half metallic ferromagnet HMF behavior (ref. [14] and therein cited works), i.e. it is metallic for majority spins (↑) and insulating for minority spins (↓) (cf. Fig. 5). The calculated magnetic band structure was confirmed later on by magneto-optic spectroscopy investigations [15].

In these contexts and the existence of several nitride-fluorides, it becomes relevant to examine in the framework of the DFT the changes of crystal, electronic and magnetic structures as well as the bonding properties in the new chromium nitride fluoride, CrNF with respect to isoelectronic $CrO_2$.

## 2  Computational details

Two computational methods within the DFT were used in a complementary manner. The Vienna a*b initio* simulation package (VASP) code [16, 17] allows geometry optimization and total energy calculations. For this we use the projector augmented wave (PAW) method [17,18], built within the generalized gradient approximation (GGA) scheme following Perdew, Burke and Ernzerhof (PBE) [19]. The conjugate-gradient algorithm [20] is used in this computational scheme to relax the atoms. The tetrahedron method with Blöchl corrections [18] as well as a Methfessel-Paxton [21] Brillouin-zone (BZ) integrals were approximated using the special k-point sampling of Monkhorst and Pack [22]. A large number of 150 k-points was used in the irreducible wedges of the respective BZ to perform total energy calculations. The optimization of the structural parameters was performed until the forces on the atoms were less than 0.02 eV/Å and all stress components less than 0.003 eV/Å$^3$. Calculations assuming firstly a non spin polarized (NSP) configuration then spin polarized (SP) calculations were carried out.

Subsequent all-electron calculations, equally based on DFT with a GGA functional were carried out with the optimized parameters for a full description of the electronic and magnetic structures and the chemical bonding properties. They were performed using the full potential scalar-relativistic augmented spherical wave (ASW) method [23]. In the ASW method, the wave function is expanded in atom-centered augmented spherical waves, which are Hankel functions and numerical solutions of Schrödinger's equation, respectively, outside and inside the so-called augmentation spheres. In the minimal ASW basis set, we chose the outermost shells to represent the valence states and the matrix elements were constructed using partial waves up to $l_{max} + 1 = 3$ for Cr and $l_{max} + 1 = 2$ for N and F. Low energy lying F($2s$) were considered as core states. In order to optimize the basis set, additional augmented spherical waves are placed at carefully selected interstitial sites (IS). Self-consistency was achieved when charge transfers and energy changes between two successive cycles were such as: $\Delta Q < 10^{-8}$ and $\Delta E < 10^{-6}$ eV, respectively. The BZ integrations were performed using the linear tetrahedron method within the irreducible wedge [18]. The calculations are carried out assuming firstly spin degenerate, non spin-polarized (NSP), then spin polarized (SP) magnetic calculations for both ferromagnetic and antiferromagnetic long range orders are done for the identification of the ground state magnetic structure of the most stable model structure. The relative magnitude of the chemical bonding is obtained based on the overlap population analysis: $S_{ij}$, i and j being two chemical species. The crystal orbital overlap population (COOP) criterion is used [24]. In the plots positive, negative and zero COOP magnitudes indicate bonding, anti-bonding and non-bonding interactions respectively.

### 3  Geometry optimization and EOS.

a- Geometry optimization

Considering the rutile structure of $CrO_2$, it is tetragonal with $P4_2/mnm$ space group (SG) N$^o$ 136, with $a$= 4.23 Å and $c$= 2.92 Å lattice parameters, d(Cr-O)~1.87 Å and atomic positions: Cr ($2a$) at (0,0,0); O($4f$) at (x,x,0; -x,-x,0; -x+½,x+ ½,½; -x+½, -x+½,½), x ~0.305 [6-10]. With two formula units (FU) per cell the stoichiometry is $Cr_2O_4$. A preliminary test was done by ordering N and F at the O ($4f$) positions with $Cr_2N_2F_2$ stoichiometry and carrying out unconstrained geometry optimization runs, the resulting symmetry becomes orthorhombic. Anion ordered rutile-type derivatives obeying group-subgroup relations were examined by Baur [25] among which there are two orthorhombic structures: $CoReO_4$ (*Cmmm* SG N°65; Z= 2 FU) and $MgUO_4$ (*Imam* SG N° 74; Z = 4 FU) –FU: formula unit–.

Both exhibit cationic and anionic ordering (cf. Table 1). We confront the two candidate structures through determining the ground state structure and deriving equilibrium values based on the energy-volume equation of state (EOS). Full analyses of the electronic and the magnetic structure as well as the properties of chemical bonding are subsequently done for the ground state structure properties.

For the purpose of geometry optimization and search for the ground state structure, non spin polarized (NSP) were carried out starting from $CoReO_4$ (based centered orthorhombic) and $MgUO_4$ (body centered orthorhombic) types [25]. The respective orthorhombic symmetries were preserved at self consistent convergence. The fully relaxed structure parameters and the corresponding energies and volumes are given in Table 1. The energies are close but point out to $MgUO_4$ –type as the ground state with $\Delta E$ = -0.165 eV/FU. The distance trends of d(Cr-F) > d(Cr-N) shown in both sets of results follow from the ionic *versus* covalent bonding and interestingly d(Cr-O) is situated in-between: d(Cr-F)~1.93-1.98 Å > d(Cr-O)~1.87 Å> d(Cr-N)~1.78-1.77 Å. The sequence of distances is found to follow inversely the sequence of electronegativity of the anions: $\chi_N$ (3.04) < $\chi_O$ (3.44) < $\chi_F$ (3.98). The resulting structures are shown in Fig. 1 highlighting the distorted *$CrN_3F_3$* polyhedra differentiated for Cr1 and Cr2 surroundings. The octahedra are edge sharing, as in rutile.

The cohesive energies for CrNF candidate structures are obtained and compared with that of $CrO_2$ by subtracting the computed atomic energy of Cr (E = -9.50 eV) and the corresponding dimmers from the total energy {E(O2): -11.21 eV, E(N2): -20.78 eV, E(F2): -3.71 eV}. Then $E_{coh.}$(CrNF *Cmmm*) = -1.045 eV/FU, $E_{coh.}$(CrNF *Imam*) = -1.107 eV/FU and $E_{coh.}$($CrO_2$ *P4/mnm*) = -1.303 eV/FU. Clearly the oxide is more cohesive than the nitride fluoride which is more cohesive in its ground state *Imam* $MgUO_4$ –derived structure. These results point out to the possibility of actually synthesizing the compound and presently investigated protocols are developed upon in the last section of the paper.

In view of the ferromagnetic ground state of $CrO_2$ with 2 $\mu_B$/FU magnetization, further geometry optimized calculations were carried out for isoelectronic CrNF considering two spin populations, i.e. spin polarized (SP) calculations. In both structural types a 2 $\mu_B$/FU was found at self consistent energy convergence with a larger volume and stabilized energy: $\Delta E$ (CrNF *Cmmm*)= -0.24 eV/FU and $\Delta E$ (CrNF *Imam*)= -0.35 eV/FU. This suggests a tetravalent character of Cr similarly to $CrO_2$. This point is discussed further while deriving the energy –volume equation of state (EOS) and the band structure in next sections.

b- Energy-volume EOS

For confronting the relative stabilities of two varieties in NSP and SP configurations, one needs establishing the energy-volume equation of state (EOS). In fact the calculated total energy pertains to the cohesive energy within the crystal because the solution of the Kohn-Sham DFT equations yields the energy with respect to infinitely separated electrons and nuclei. In as far as the zero of energy depends on the choice of the potentials, somehow it becomes arbitrary meaning it is shifted but not scaled. However the energy derivatives as well as the EOS remain unaltered. For this reason one needs to establish the EOS and extract the fit parameters for an assessment of the equilibrium values. This is done from (E,V) set of calculations around minima found from geometry optimization. The resulting E = $f$(V) curves are shown in Fig. 2. They have a quadratic variation which can be fitted with energy-volume Birch-Murnaghan EOS to the 3$^{rd}$ order [26]:

$$E(V) = E_o(V_o) + [9/8]V_oB_o[[(V_o)/V)]^{[2/3]}-1]^2 + [9/16]B_o(B'-4)V_o[[(V_o)/V)]^{[2/3]}-1]^3,$$

where $E_o$, $V_o$, $B_o$ and $B'$ are the equilibrium energy, the volume, the bulk modulus and its pressure derivative, respectively. The fit results in the inserts of Fig. 2 show the trend of smaller volumes and higher energies of the NSP forms versus SP pointing to stable magnetically polarized compounds. The energies are such that $\Delta E_{SP-NSP}$(CrNF *Cmmm*)= -0.29 eV/FU and $\Delta E_{SP-NSP}$(CrNF *Imam*)= -0.35 eV/FU. Interesting features appear for the zero pressure bulk modules which are systematically higher for the NSP configuration as with respect to SP as one would expect from the larger volume for the latter due to the magnetic (negative) pressure. At large volume the separation between the SP and NSP E(V) curves is maximum while they meet at low volume. This is expected in as far as the effect of decreasing volume (increasing pressure) is to reduce the localization of the *d* states needed for the onset of the magnetic moment. Consequently the magnetization should vanish at high pressure, i.e. at volumes below the equilibrium. This can be estimated from the Birch relationship providing the pressure [26]:

$$P = (B_o/B') [(V_o/V')^{B'}-1].$$

Taking the fit values $B_o$, $B'$, $V_o$ of the SP curve for the $MgUO_4$–derived structure and the volume V' as V(NSP), P is calculated to be 11 GPa. This value is much smaller than the one calculated for $CrO_2$ of 120 GPa [27]. CrNF is then assigned a soft magnetic behavior while $CrO_2$ is a hard ferromagnet. The plots of the change of the magnetization with the volume are shown for $MgUO_4$–type CrNF and for $CrO_2$ in Fig. 3. The magnetization increases with volume and reaches saturation with 8 $\mu_B$/cell for CrNF and 4 $\mu_B$/cell for $CrO_2$. What need to be highlighted in the plots are the different behaviors of the two compounds around the equilibrium volume $V_0$ indicated by a vertical line: in CrNF there is a nearly immediate decrease of the magnetization below $V_0$ whereas in $CrO_2$ the saturation magnetization is kept far below $V_0$. This nicely reflects the pressure magnitudes above.

## 4 All-electrons calculations.

The electronic and magnetic structure and the chemical bonding properties were calculated for the ground state structure of CrNF with the optimized lattice parameters. Calculations with the $CoReO_4$-type CrNF led to similar results.

Firstly NSP calculations were done to assign a role of each chemical species in the valence band (VB) and conduction band (CB). With the ASW method one starts from neutral atoms; at self consistency (energy and charge) using 432 k-points generated from 1728 (12 × 12 × 12) k-points in the orthorhombic BZ, there is charge redistribution between the atomic species. This does not translate the ionic picture as $F^-$ despite the large ionic character of fluorine but the charge transfer is actually observed from Cr to N, F and IS, in agreement with the expectations of chemistry, i.e. from the metallic species to the anions (N, F) and IS which receive charge residues, i.e. less than 0.1 electron. Further the band structure will be shown to provide an illustration for the expected chemical picture of the nitride-fluoride.

**4.1 Non magnetic calculations.**

The NSP site projected DOS (PDOS) accounting for site multiplicity are shown in Fig. 4. The energy reference along *x* is at the Fermi level $E_F$ crossing a large PDOS intensity arising from Cr *d* states which are mainly centered in the empty conduction band, above $E_F$. For the sake of clarity we do not show small contribution PDOS arising from IS. The lower part of the valence band (VB) is dominated by N(2s) at ~ -14 eV then F(2p) in the range {- 10, -7 eV} and N(2*p*) in the range {-5, -1 eV}. The PDOS below the two *p* blocks point to a larger Cr-N mixing than Cr-F. This is further made explicit in the analysis of the chemical bond in next section. Besides the mixing of states the NSP DOS allow assessing the instability of the compound from the large Cr PDOS at $E_F$ in such a non magnetic configuration. The mean field Stoner theory of band ferromagnetism [28] can be applied to address the tendency for spin polarization. The total energy of the spin system results from the exchange and kinetic energies. Referring the total energy to the non-magnetic state, this is expressed as: E = *constant*{1- $\mathcal{J}n(E_F)$}. In this expression, $\mathcal{J}$ (eV) is the Stoner integral, which is calculated and tabulated for the metals [29] and $n(E_F)$ (1/eV) is the PDOS value for a given state -mainly *d*- at the Fermi level in the non-magnetic state. If the unit-less Stoner product $\mathcal{J}n(E_F)$ is larger than 1, E is lowered and the system stabilizes in a magnetically ordered configuration. Then the product $\mathcal{J}n(E_F)$ provides a criterion for the stability of the spin system. From [29] the value of $\mathcal{J}${Cr(3*d*)} is 0.038 eV. With $n(E_F)$ value of 9 eV$^{-1}$, the calculated $\mathcal{J}.n(E_F)$ is 3.4 and the Stoner criterion 1- $\mathcal{J}n(E_F)$ is largely negative leading to energy lowering upon the onset of magnetization, *i.e.* intra-band spin-polarization should occur when spin polarization is allowed. Subsequent spin polarized (SP) calculations with equal initial ↑ and ↓ spin populations, were carried out. At self-consistency a finite magnetization can be identified within an implicit long range ferromagnetic order. Nevertheless if magnetic exchange energy is not sufficient, zero local moments can result, so that the calculations are not biased initially. This also depends on of the BZ mesh precision, i.e. the calculations are usually carried out in steps of increasing precision until no more changes are observed in variational energy and magnetic moments.

The lower panel of Fig. 4 shows the bonding characteristics based on the analysis of the overlap populations $S_{ij}$ with the COOP criterion. The bonding within the valence band VB is mainly between chromium on one hand and N and F on the other hand. It is differentiated between the two chromium sites with Cr1-F at -10 eV, Cr2-N at -4 eV. Note the low magnitude Cr2-F COOP versus Cr1-N. The overall prevalence of Cr-N over Cr-F follows the shorter Cr-N versus Cr-F distances and signals the rather covalent nature of the nitride fluoride. The itinerant Cr-*d* states mix with the *p*-states of N and F following their PDOS positions in the DOS panel. The bonding between the two anionic sublattices is of antibonding nature because of the involvement of their states in the bonding with Cr. The large antibonding Cr1-F $E_F$ COOP's are due to the near saturation of F-2*p* i.e. close to 2$p^6$ so that the electrons brought by the Cr-F bond go into anti-bonding COOP's. This is opposed to the Cr-N COOP's at $E_F$ which have much less intensity. The large anti-bonding COOP's at $E_F$ is another signature of the instability of the nitride fluoride in such a spin degenerate.

**4.2 Ferromagnetic state and search for magnetic ground state.**

Upon charge and energy self-consistent convergence the total variational energy decreases largely with respect to the NSP configuration by $\Delta E(SP - NSP)$= -1.41 eV/cell. This gain of energy is due to the magnetic exchange leading to the onset of a magnetic moment as it is shown below. The total magnetization is 8.0 $\mu_B$/cell, i.e. 2.0 $\mu_B$/FU with different local spin moments distributions: M(Cr1) = 2.183 $\mu_B$, M(Cr2) = 2.317 $\mu_B$; M(F) = 0.025 $\mu_B$ and M(N) = -0.348 $\mu_B$ with the remaining contributions from IS (interstitial part). The negative moment on nitrogen is of induced nature and arises from the Cr-N strong hybridizing show in the preceding section.

The result of an integer total magnetization is the result of the contribution of all the chemical constituents, not only Cr. It is illustrated with the plot of the SP PDOS shown in Fig. 5. The magnetic exchange causes the majority spins to shift down in energy and the minority spins to shift up in energy, proportionally to the magnitude of the developed magnetic moment. This is actually observed for Cr $d$ states and less for F whereas the opposite is observed for $s$ and $p$ states of nitrogen which possesses a negative moment. The striking feature is observed at the Fermi level with large PDOS for ↑ spin mainly due to Cr1 and Cr2 and a gap opening of ~0.5 eV for ↓ spin. Comparing the SP DOS with those formerly obtained for $CrO_2$ in lower panel of Fig. 5 [14]; low energy lying O(2s) are not shown. The DOS present similarities as to the metallic ↑ spin and the gap opening at ↓ spin. However the gap magnitude of ~2 eV is much larger than the band gap in CrNF; this arises from the larger covalent overall Cr-N bonding versus Cr-O and also contribute explaining the different magnetic responses of the two compounds to pressure as discussed above.

Lastly for a search of the magnetic ground state, the $Cr1_2Cr2_2N_4F_4$ stoichiometry in the $MgUO_4$ –type structure (Table 1) allows splitting the unit cell into two magnetic subcells with one $Cr1Cr2N_2F_2$ subcell as (UP SPINS) and the other $Cr1Cr2N_2F_2$ subcell as (DOWN SPINS), thus enforcing the antiferromagnetic order. This is especially interesting in as far as CrN is an antiferromagnet in its orthorhombic structure (cf. [30] and therein refs.). At self energy convergence there is full compensation of spins and an energy difference ΔE(SP-F – SP-AF) = -0.13 eV/FU. Then the long range order of the ground state is ferromagnetic.

## 5  Prospective synthetic routes.

In view of the favorable cohesive energies of CrNF in both rutile derived model structures and the identified close relationship with technologically important $CrO_2$. Synthetic routes are presently being evaluated along two different protocols:

    a- the ammonolysis of chromium fluoride precursor such as the protocol used on $ZrF_4$ for obtaining ZrNF [31];

    b- the method using fluorinated fluxes on CrN following a protocol proposed for preparing ZrNF [5].

The latter protocol seems relevant from the electronic structure characteristics of CrNF, whereby within CrN, covalent Cr-N bonding is expected to be modified but still keep prevailing upon introducing fluorine either by $F_2$ fluxes or by use of $F_2$ pressure [32].

**Acknowledgements:**

Discussions and exchange on the topic of nitride-fluorides and on the synthesis routes with Prof. Gérard Demazeau, Dr Alain Largeteau and Prof. Jean Etourneau (France), Prof. M.A. Subramanian (USA) are gratefully acknowledged.

Figures

Fig.1 : CrNF in $CoReO_4$ (top) and $MgUO_4$ (bottom) ordered rutile derivatives highlighting the edge sharing mixed anion (N,F) octahedra at Cr1 (grey) and Cr2 (yellow) sites.

Fig.2 Energy-volume curves for CrNF in $CoReO_4$ (top) and $MgUO_4$ (bottom) types in spin degenerate (NSP) and spin-polarized (SP) configurations and fit parameters from Birch-Murnaghan equations of states.

Fig.3 Total magnetization volume change for CrNF in $MgUO_4$-type ground state structure (top) and in $CrO_2$ (bottom). Connecting line are shown as a guide for the eye.

Fig.4 CrNF with ground state $MgUO_4$ structure: Spin degenerate site projected DOS (top) and chemical bonding (COOP).

Fig. 5 Site and spin projected density of states of CrNF (top) and $CrO_2$ (bottom).

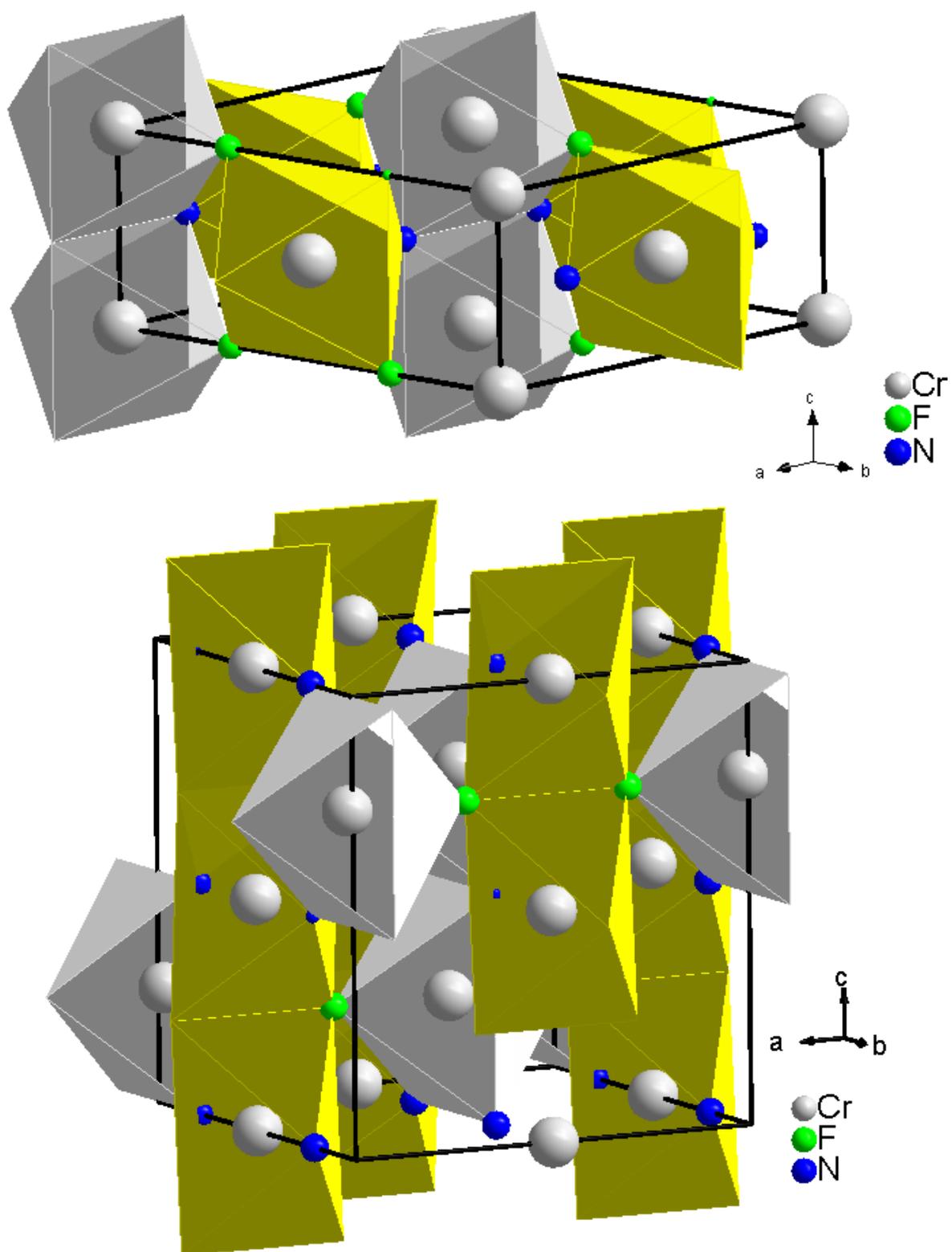

FIG. 1

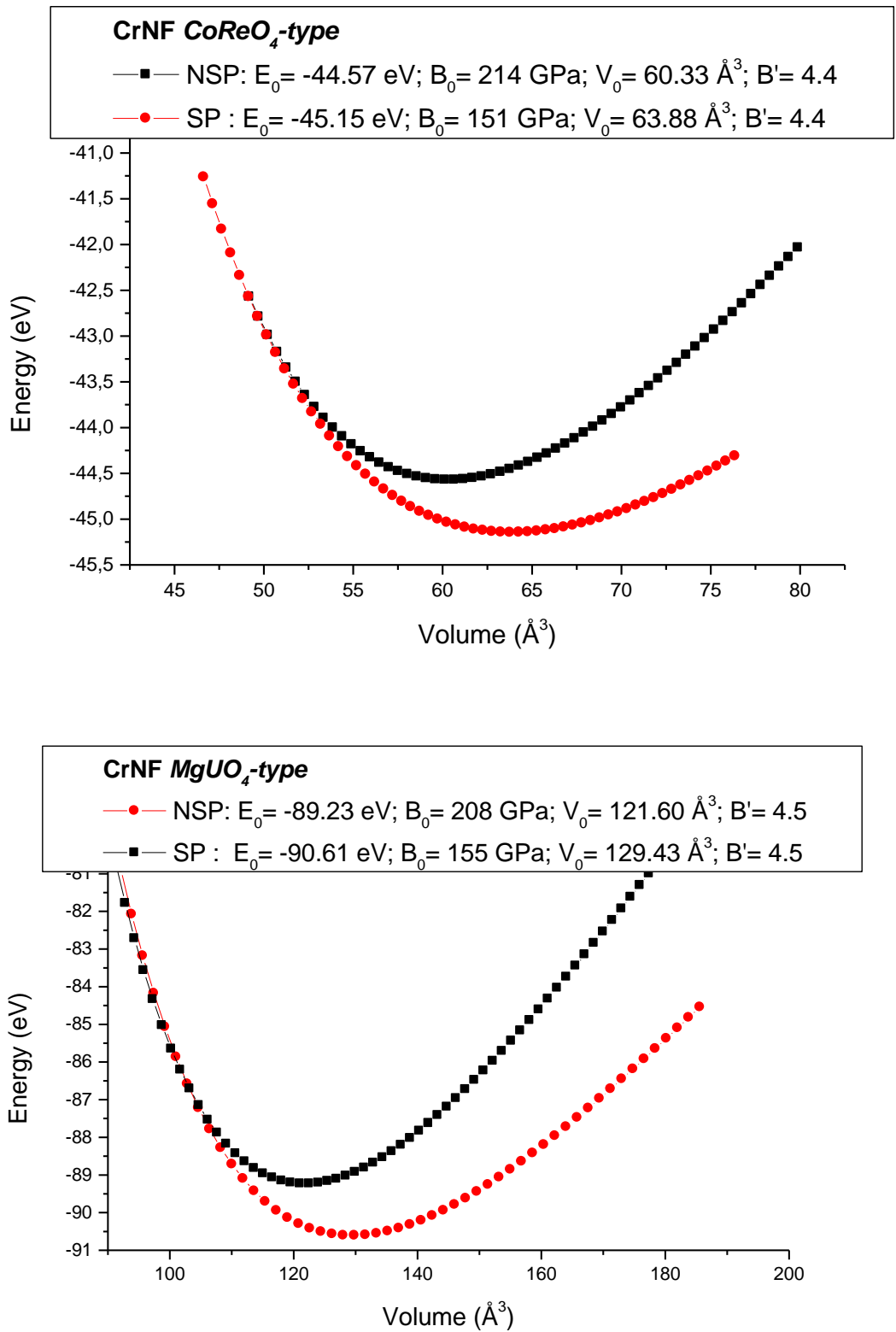

FIG. 2

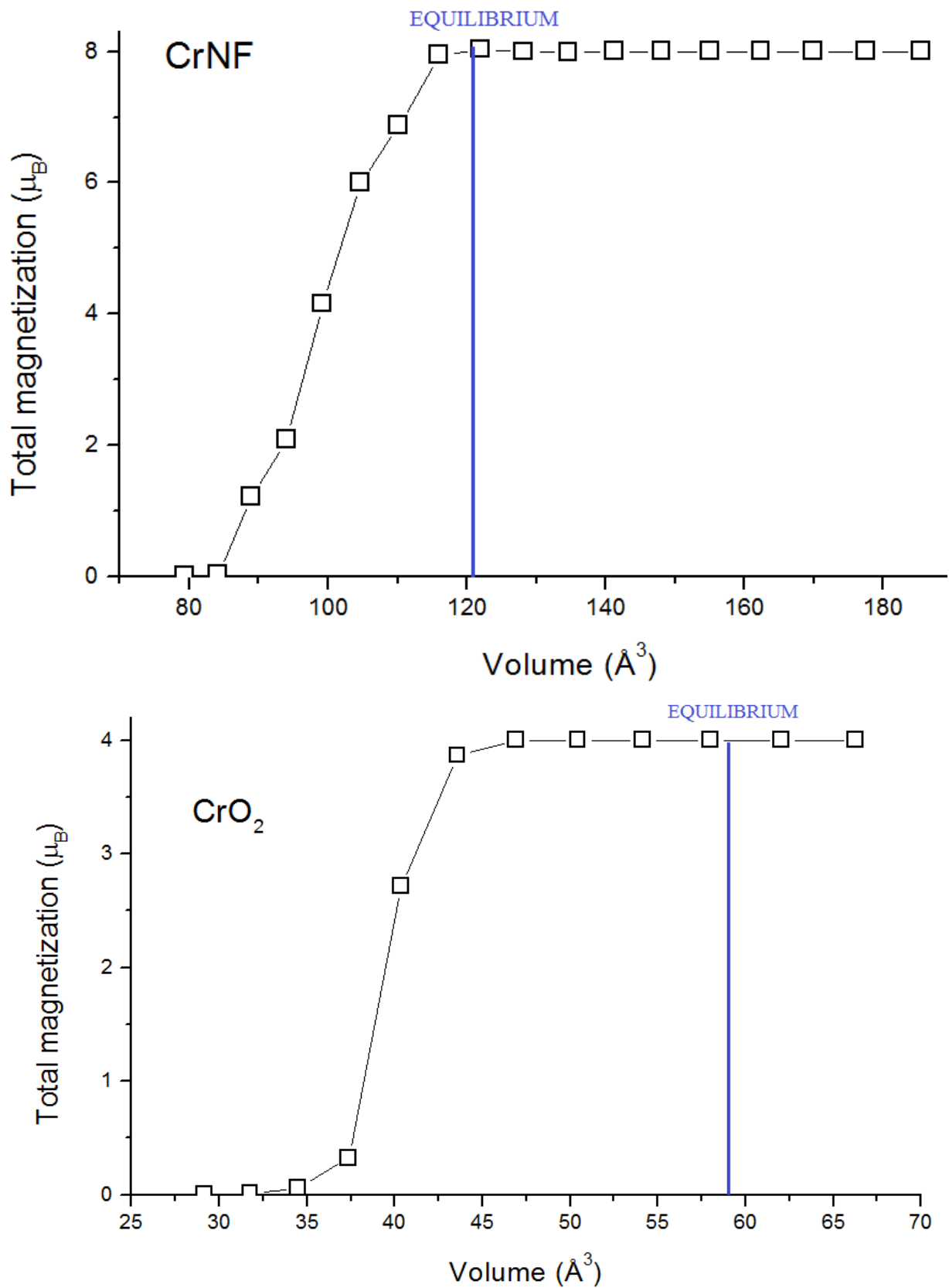

FIG. 3

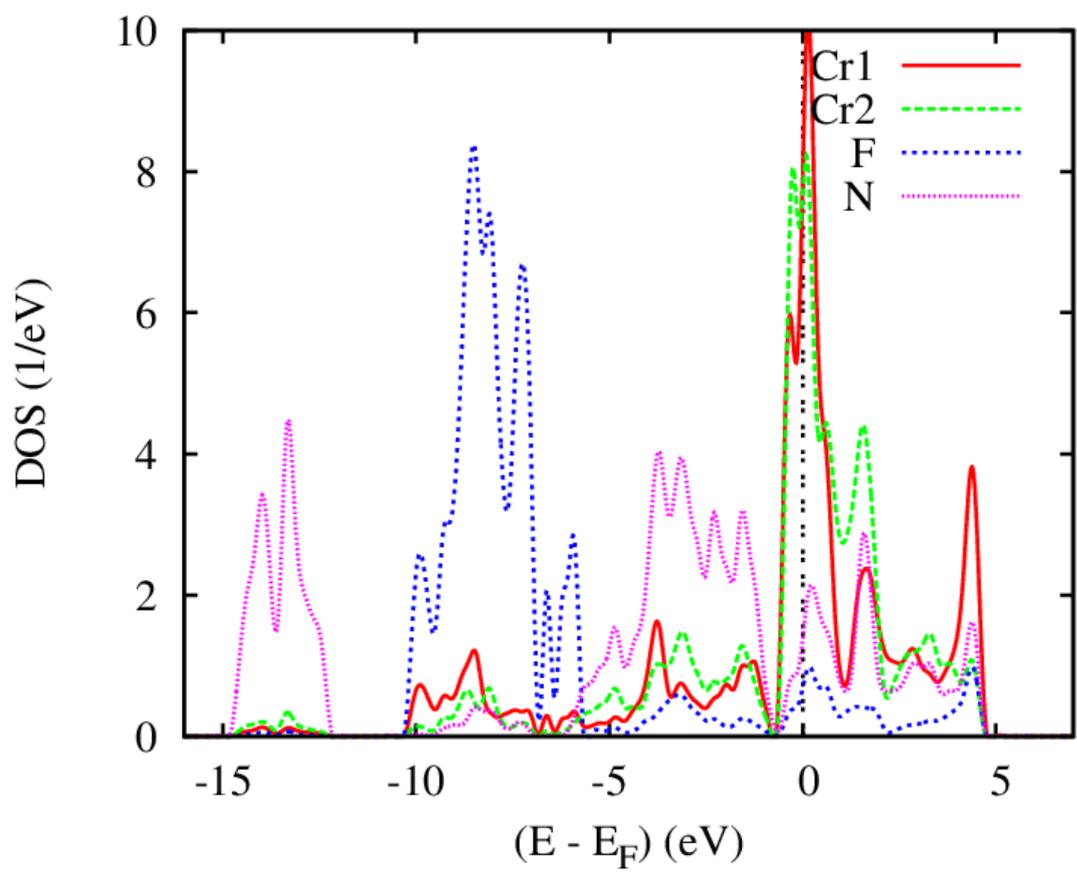

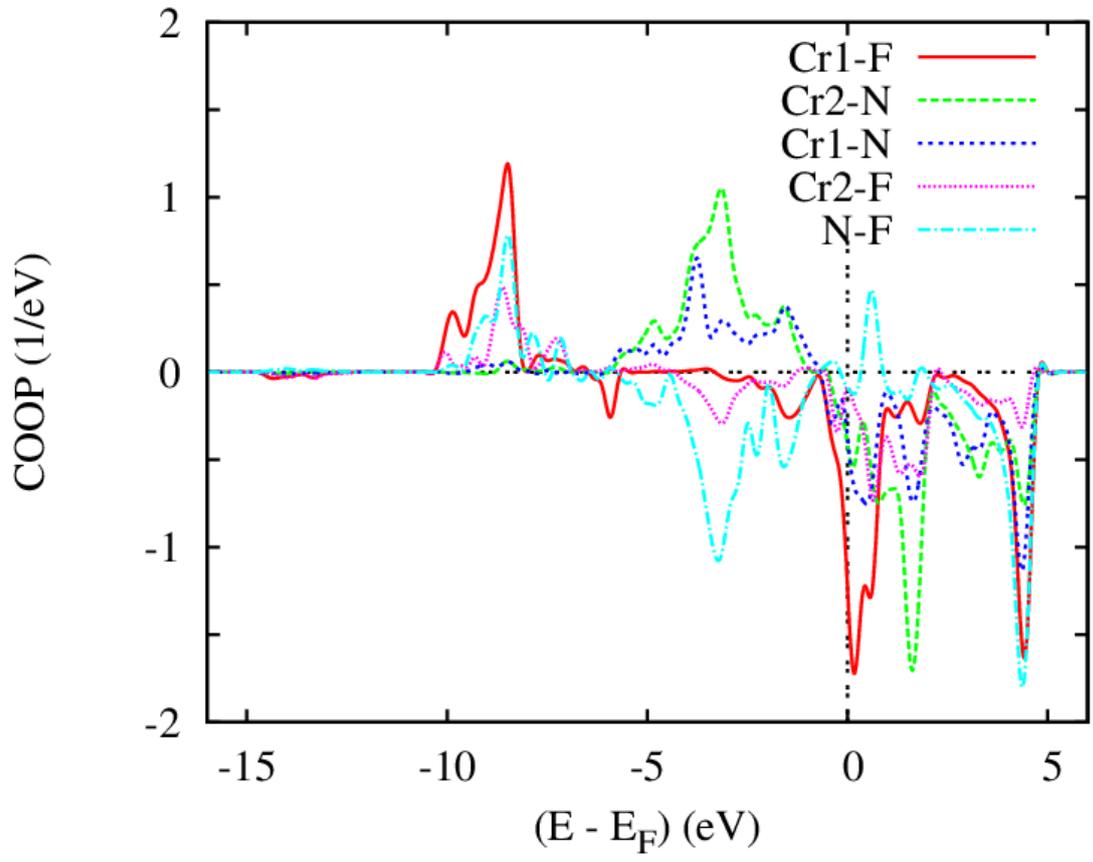

FIG. 4

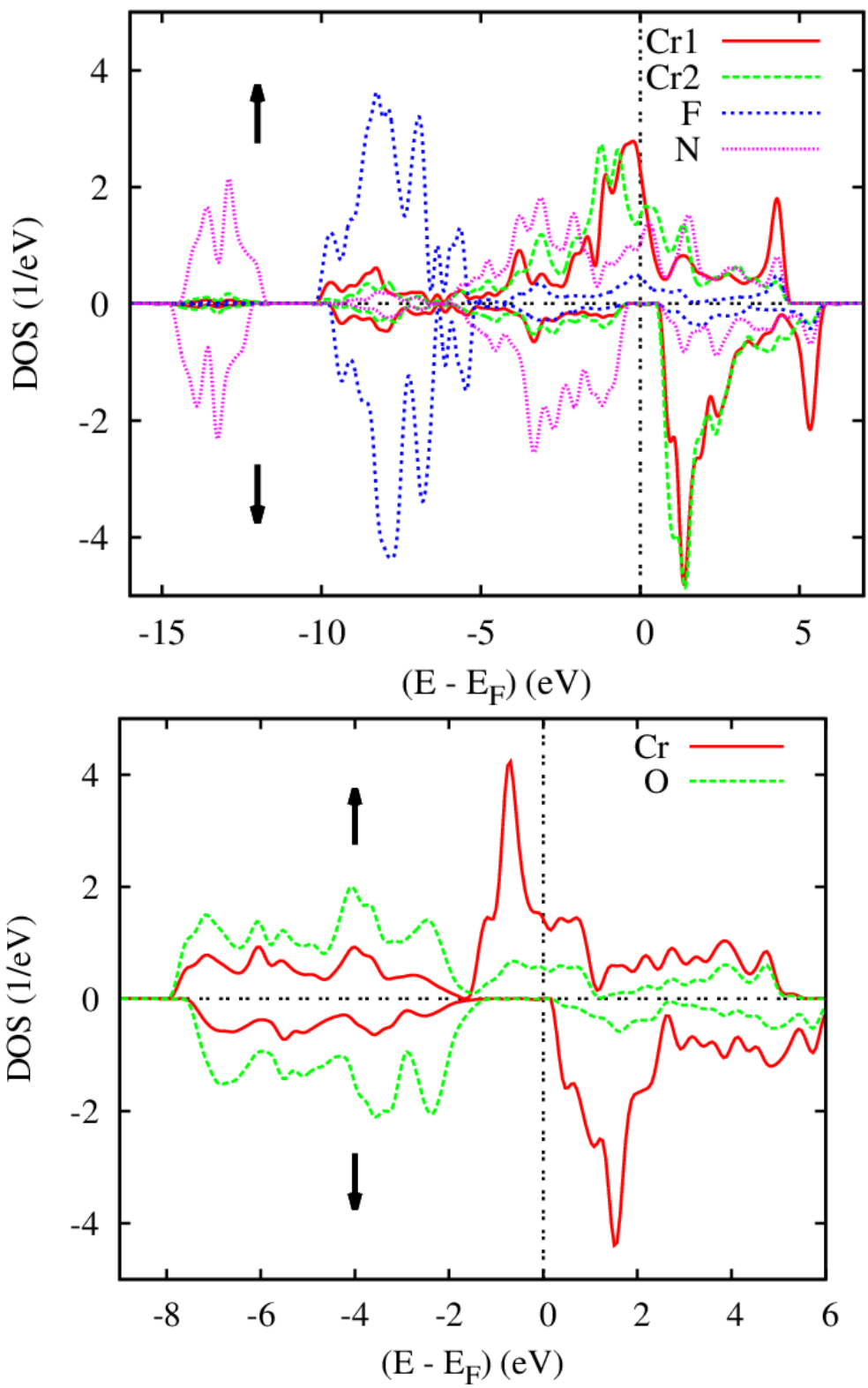

FIG. 5